\PassOptionsToPackage{hyphens}{url}
\RequirePackage{fix-cm}
\documentclass[sn-mathphys,Numbered]{sn-jnl}

\usepackage{mathrsfs}%
\usepackage[title]{appendix}%
\usepackage{textcomp}%
\usepackage{manyfoot}%
\usepackage{algorithmicx}%
\usepackage{algpseudocode}%

\usepackage{framed}
\usepackage{seqsplit}
\usepackage{float}
\usepackage{subfigure}
\usepackage{listings,amsfonts}
\usepackage{amsmath,pifont}
\usepackage{xspace}
\usepackage{bbding}
\usepackage{balance}
\usepackage{hyperref,endnotes}
\usepackage{booktabs}
\usepackage{array}
\usepackage{algorithm}
\usepackage{multirow,makecell}
\usepackage{enumitem}
\usepackage[normalem]{ulem}
\usepackage{graphicx}
\usepackage{amsthm}
\usepackage{hhline} 
\usepackage{xcolor}
\usepackage{stfloats}
\usepackage[justification=centering]{caption}

\usepackage{algpseudocode}
\usepackage[normalem]{ulem}
\usepackage{shortcuts}

\usepackage{amssymb}

\begin{document}

\title{The Devil Behind the Mirror: Tracking the Campaigns of Cryptocurrency Abuses on the Dark Web} 

\author[1]{\fnm{Pengcheng} \sur{Xia}}

\author[1]{\fnm{Zhou} \sur{Yu}}

\author[2]{\fnm{Kailong} \sur{Wang}}

\author[2]{\fnm{Kai} \sur{Ma}}

\author[1]{\fnm{Shuo} \sur{Chen}}

\author[3]{\fnm{Xiapu} \sur{Luo}}

\author[4]{\fnm{Yajin} \sur{Zhou}}

\author[4]{\fnm{Lei} \sur{Wu}}

\author[5]{\fnm{Guangdong} \sur{Bai}}

\affil*[1]{\orgname{Beijing University of Posts and Telecommunications}, \city{Beijing}, \country{China}}

\affil[2]{\orgname{Huazhong University of Science and Technology}, \city{Wuhan}, \country{China}}

\affil[3]{\orgname{The Hong Kong Polytechnic University}, \city{Hong Kong}, \country{China}}

\affil[4]{\orgname{Zhejiang University}, \city{Hangzhou}, \country{China}}

\affil[5]{\orgname{University of Queensland}, \city{Brisbane}, \country{Australia}}

\abstract{
  The dark web has emerged as the state-of-the-art solution for enhanced anonymity. Just like a double-edged sword, it also inadvertently becomes the safety net and breeding ground for illicit activities. Among them, cryptocurrencies have been prevalently abused to receive illicit income while evading regulations. Despite the continuing efforts to combat illicit activities, there is still a lack of an in-depth understanding regarding the characteristics and dynamics of cryptocurrency abuses on the dark web. 
  
  In this work, we conduct a multi-dimensional and systematic study to track cryptocurrency-related illicit activities and campaigns on the dark web. We first harvest a dataset of 4,923 cryptocurrency-related onion sites with over 130K pages. Then, we detect and extract the illicit blockch\-ain transactions to characterize the cryptocurrency abuses, targeting features from single/clustered addresses and illicit campaigns. Throughout our study, we have identified 2,564 illicit sites with 1,189 illicit blockchain addresses, which account for 90.8 BTC in revenue. Based on their inner connections, we further identify 66 campaigns behind them. Our exploration suggests that illicit activities on the dark web have strong correlations, which can guide us to identify new illicit blockchain addresses and onions, and raise alarms at the early stage of their deployment.
}
 

\maketitle

\section{Introduction}
\label{sec:intro}
Echoing the increasing need to protect users' online browsing privacy, the dark web has emerged as a state-of-the-art solution which utilizes a specifically configured and customized communication protocol to conceal a user's browsing footprints. In particular, most of the dark web contents~\cite{darkweb_news} are hosted by the Tor Hidden Services (or onion services) that are only accessible through Tor networks. We thus target onion services in this paper, and refer to them as the dark web hereafter for simplicity. 

Accompanied by such enhanced user anonymity, the dark web also inadvertently serves as a safety net and breeding ground for illicit activities. Typical examples include fraud, ransomware, trafficking of illegal items (e.g., personal information, counterfeit goods, drugs, guns, and even humans), and illegal services (e.g., murder for hire and child pornography). As a consequence, recent years have witnessed an explosive surge of crimes related to the dark web. For example, the leading dark web market named AlphaBay~\cite{AlphaBay} alone has served over 200,000 users with over \$1 billion generated transactions from 2014 to 2017. Another example provided in the Cyber Threat Report 2023~\cite{cybersec_report} by the National Cyber Security Centre highlights that the dark web has long harbored a flourishing market for stolen personal information and credentials, including logins and passwords. 

Due to the capability of obfuscating the cross-border money flows while retaining the owner's anonymous identity, cryptocurrencies are deemed an ideal means of payment or an investment fraud currency~\cite{Europol_Spotlight} that can effectively evade monetary regulations around the world. Consequently, the abuse of cryptocurrencies has been found prevalent among dark web services. According to the cybercrime intelligence company Recorded Future~\cite{recorded_future}, Bitcoin is accepted at almost all e-commerce storefronts on the dark web~\cite{panda2019growth}. As reported by a study from Chainalysis~\cite{chainanalysis}, record revenue of \$1.5 billion worth of cryptocurrencies has been brought into the dark web marketplaces in 2022. 

The continued proliferation of cybercrimes on the dark web has drawn the attention of law enforcement. In April 2022, Hydra Market, which was considered the world's largest dark web marketplace, was partially taken down by a campaign cooperated between the US and German authorities~\cite{hydra_market}. Prior to this incident, a series of high-profile dark web marketplaces were shut down, such as Silk Road, AlphaBay, and Hansa. They merely expose the tip of an iceberg, as large numbers of newly emerged illicit activities would quickly refill the void created by the takedowns without being promptly detected and tracked. 

Given the characteristics of the dark web and cryptocurrencies, detecting and tracking illicit activities can be challenging. 
From the perspective of the dark web, the key challenge is to effectively identify the dark web sites and further determine if they are related to illicit activities. Despite the relevant efforts~\cite{He2019ICISS,Khan2019} in the literature, there is still a lack of an automatic method to detect and classify illicit dark web sites on a fine-grained level. 
From the perspective of cryptocurrencies, the key challenge arises from the lack of an in-depth understanding of the transaction-related and address-related characteristics in the specific domain of dark web illicit activities. Some efforts~\cite{lee2019cybercriminal,kanemura2019identification,dearden2023follow} conduct studies on cryptocurrency-related transactions in the dark web. However, no existing studies have investigated the communities behind them, which could reflect the properties of illicit campaigns beyond the characteristics of individual illicit sites or addresses. 

\noindent \textbf{Our Work.} In this work, we aim to propose a systematic approach to effectively identify illicit dark websites, track the cryptocurrency-related illicit activities, reveal the illicit campaigns behind them, and further characterize the cryptocurrency abuses. We first implement a dark web crawler and deploy it to collect a dataset containing 4,923 onion sites with over 130K pages that are related to cryptocurrencies. Based on this dataset, we manually assemble a ground truth dataset, and create a reliable classifier to detect and categorize illicit onion sites into 12 fine-grained types~(Section~\ref{sec:study-design}), with 2,564 illicit sites in total. We then strive to understand their illicit activities in the dark web ecosystem from multiple perspectives, leveraging the investigation on identified blockchain addresses. In particular, we characterize the features associated with single addresses (e.g., illicit income, active period, and shared sites, detailed in Section~\ref{sec:overview}), clusters of addresses~(detailed in Section~\ref{sec:cluster}) and addresses related to illicit campaigns~(detailed in Section~\ref{sec:campaign}). 

\noindent \textbf{Key Findings.} Our study has revealed previously unknown features regarding illicit campaigns, as highlighted below. We defer the details to our evaluation sections~(Section~\ref{sec:overview},~\ref{sec:cluster} and~\ref{sec:campaign}). 

\begin{itemize}
    \item\textbf{Imbalanced distribution of illicit income.} We find that the majority~(56\%) of total identified illicit income concentrates over a small number~(30 addresses, 7\% of total addresses) of long-living addresses~(i.e., more than three years). This suggests their prolonged and covert illicit activities. Meanwhile, most addresses~(63\%) only exist for less than a year. 

    \item \textbf{Multi-faced campaigns.} We reveal that illicit campaigns tend to host multiple onion sites to launch various malicious activities. 26 out of the 66 identified campaigns host more than ten onion sites. Despite such diversity, more than half of the campaigns maintain a single blockchain address for receiving payments. 
    
    \item \textbf{The darkness on the web.} Among the prevalent illicit activities on the dark web, scam is a prominent type. We identified 39\% of the campaigns targeting investment scams, which account for a quarter of the total illicit income. They typically utilize different disguises to hide their malicious purpose, more to be detailed in Section~\ref{subsec:abuse_rep}.

\end{itemize}

\noindent \textbf{Contributions.} In summary, this work mainly contributes to the following aspects:

\begin{itemize}
    \item \textbf{A systematic study.} We systematically investigate the illicit activities
    among dark web sites, from the special angle of cryptocurrency-related abuses. Our study unveils richer connections among illicit sites and larger numbers of illicit campaigns underlying the dark web ecosystem, as compared to the existing works in the literature.
    
    \item \textbf{Characterizing cryptocurrency abuses.} We conduct a multi-dimensional evaluation of the cryptocurrency abuses, focusing on the characteristics traceable via the blockchain addresses. We show that most illicit activities have strong connections that can be grouped into illicit campaigns. 
    
    \item \textbf{Practical Results.} We have collected a dataset containing 4,923 onion sites with over 130K pages that are related to cryptocurrencies. Among them, we identify 2,564 illicit sites with 8,138 blockchain addresses. Regarding the cryptocurrency abuses, we have identified 1,189 illicit addresses with a total transaction volume of 90.8 BTC. These illicit sites are further grouped into 66 campaigns, and some of them are both active in the dark and surface web.
  
\end{itemize}

\section{Backgrounds}\label{sec:background}

\subsection{The Dark Web} The dark web, different from the surface web available to the general public, requires specific software, tools, and authorization to access. It is a less accessible subset of the deep web, the web not indexed by regular search engines. The dark web relies on network connections made between trusted peers and sophisticated encryption technologies to enhance the anonymity of users, and one of the most famous networks is The Onion Router (Tor) network~\cite{tor}. 

\subsection{The Onion Services} Tor uses the onion service, which encrypts and routes traffic through multiple servers around the world, to hide the users' private information and Internet activities. An onion website can be accessed through a domain name with a top-level domain (TLD) suffix \texttt{.onion} (e.g., \texttt{edx2f26lcagct5p\-o.onion}). Different from domain names on the surface web, the \texttt{.onion} domain names are not acquired by registration, but are hashes of public keys. Besides, \texttt{.onion} domain names are relatively hard to find on search engines, and the most direct way to find pages is to receive a link to the page from someone who already knows it. Based on this prior knowledge, we implement a crawler to collect \texttt{.onion} sites in Section~\ref{sec:collect}. When accessing an onion site, the browser asks for the onion service directory instead of looking up an IP address through DNS. Although intended for privacy concerns, the hidden nature of the Tor network is exploited in more and more illicit activities, including selling or distributing malicious content like drugs, weapons, stolen data, malware, etc. Combined with anonymous payment systems like cryptocurrencies, the illicit activities on the Tor network become less traceable. 

\subsection{Cryptocurrency} Cryptocurrency is a kind of digital asset that takes advantage of cryptography to ensure the security of its creations and transactions. It relies on the blockchain, a shared, immutable, and distributed ledger, to record transactions and track assets in a peer-to-peer (P2P) network. The first and most well-known cryptocurrency, Bitcoin, was released in 2009. Since its debut, thousands of cryptocurrencies have emerged. According to CoinmarketCap~\cite{coinmarketcap}, there are 19K cryptocurrencies now. Considering that almost all the vendors~(91\%) in dark marketplaces accept Bitcoin as the preferred method of payment~\cite{hollandfintech}, we thus mainly focus on Bitcoin in this work. 
The core of Bitcoin is data with ownership assigned. The ownership changes when transactions take place. Bitcoin wallets keep a private key to sign transactions, providing mathematical proof that they have come from the owner of the wallet. All transactions are broadcast to the network through the mining process, where miners compete to solve the complex hash puzzle to verify transactions for Bitcoin rewards. 

\section{Tracking Cryptocurrency Abuses on the Dark Web} 
\label{sec:study-design}
In this section, we present our approach to identifying the illicit onion sites and blockchain addresses, focusing on the relevant cryptocurrency abuses on the dark web. To facilitate understanding, we first outline the steps included in the approach before describing each step in more detail. 

\begin{figure*}[t]
  \centering
  \includegraphics[width=\linewidth]{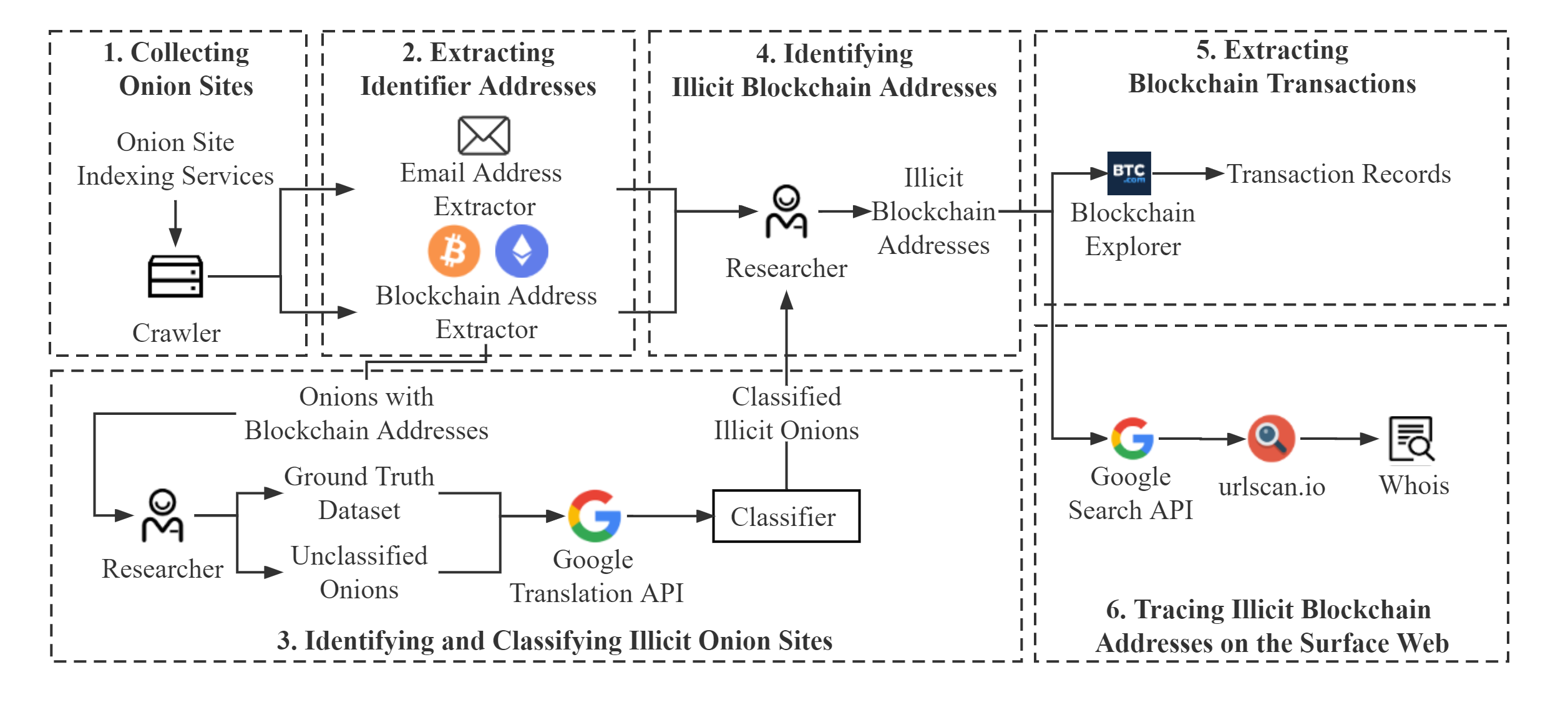}
  \caption{The Overall Workflow of Tracking Cryptocurrency Abuses on the Dark Web.} 
  \label{fig:workflow}
\end{figure*}

\subsection{Approach Overview} 
As shown in Figure~\ref{fig:workflow}, our approach consists of 6 modules, covering steps from onion site collection to illicit blockchain address tracing, as follows: 
\begin{itemize}
    \item \textbf{Step 1} Deploying a dark web crawler to identify and collect available onion sites
    \item \textbf{Step 2} Extracting blockchain and identifier (i.e., email) addresses from the collected sites
    \item \textbf{Step 3} Identifying and classifying illicit sites based on a three-phase similarity-based method
    \item \textbf{Step 4} Harvesting illicit blockchain addresses from illicit onion sites
    \item \textbf{Step 5} Extracting transaction records from illicit blockchain addresses
    \item \textbf{Step 6} Tracking illicit blockchain addresses from the surface web
\end{itemize}

\subsection{Collecting Onion Sites}
\label{sec:collect}
We utilize web scraping techniques to collect onion sites for our analysis. Considering that onion sites can only be accessed through their specific onion domain names, we resort to a repetitive crawling strategy to allow a comprehensive discovery of such names. In particular, we collect onion names as the seed for the crawler from the following two data sources. One source is surface websites providing onion addresses. The other one is the dark web indexing services, i.e., Ahmia~\cite{ahmia} and Torch~\cite{torch}. Using the collected onion names, our crawler instance visits the onion sites and records their pages. External links with a ``.onion'' suffix will be extracted from each crawled page. They will be added to the onion name queue for crawling if they have not been visited before. The crawler instance repeats the above steps until no new onion sites can be found.

Our crawler code is written in Python and uses Scrapy~\cite{scrapy}, a lightweight and open-source web crawling framework, in the implementation. We note that Scrapy only downloads the page source code during crawling, without loading or rendering the dynamic contents. Fortunately, most of the dark web pages are simple, with merely static content. Our crawler can retain sufficient reliability in this process. As further confirmation, we manually traversed and examined 200 randomly selected onion sites, and found no dynamically loaded web contents. 
We deploy our crawler instance on a remote Ubuntu server, with 4GB RAM and 80GB storage. To enable access to onion sites, our crawler is installed with Tor service. By study time, the crawler has collected $18,014$ onion sites with $394,459$ web pages in total. 
Compared with~\cite{lee2019cybercriminal} that detected over 36K onion sites, we focus on capturing the colluding malicious campaigns behind clusters of sites, which is plausible for our relatively smaller dataset. In parallel to the instability~(e.g., service offline) of the active onion sites at all times, this number difference is also possibly coupled with the active take-downs of the dark web marketplaces in recent years. 

\subsection{Extracting Identifier Addresses from Collected Onions} 
From the collected onion sites, we extract two types of address information, including blockchain addresses and email addresses. This is based on the consideration that perpetrators in the dark web use their blockchain addresses to receive cryptocurrencies, from which we can further track their identities and illicit activities. 
In addition, email addresses that are typically provided on illicit onion sites for customer service purposes can also serve as an indicator to help identify the site owner and campaigns behind the anonymity network.

\subsubsection{Extracting Blockchain Addresses}
Blockchain addresses are intrinsically hash values, the length of which varies according to their types. The blockchain addresses with the largest transaction volumes~\cite{kulal2021followness}, i.e., those for Bitcoin and Ethereum in particular, are taken into account. To effectively identify Bitcoin and Ethereum addresses, 
we use regex expressions to extract matching blockchain addresses from the onion pages, as shown below. We further use the checksum algorithm~(for Bitcoin addresses) and web3.js~\cite{webjs} API~(for Ethereum addresses) to verify the address validity separately. The extractor ends up with $15,752$ blockchain addresses ($15,450$ for Bitcoin, $302$ for Ethereum) from $4,923$ onion sites. We further filtered out addresses that are unrelated to illicit site owners, as detailed in Section~\ref{sec:illicit-addr}. 

\begin{displaymath}
Bitcoin: ([0-9a-zA-Z]\{25,39\})
\end{displaymath}
\begin{displaymath}
Ethereum: ((0x)?[0-9a-fA-F]\{40\})
\end{displaymath}

\subsubsection{Extracting Email Addresses}
An email address consists of a username, an @ symbol, and a domain name. We extract email addresses by regex expressions and eliminate those with invalid email domain names. As a result, $1,246$ Email addresses are derived from $2,736$ crawled onion sites. 

\subsection{Identifying and Classifying Illicit Onion Sites}
\label{sec:classify}
We design a three-phase similarity-based method to identify illicit onion sites and classify them according to the types of illicit activities. In brief, we first construct a small manually labeled benchmark dataset. Based on it, we use a similarity-based method to identify illicit sites that are highly similar to them. Next, we propose a TF-IDF-based method to classify the remaining sites based on text features. As our main concern is on the cryptocurrency-related abuses, we focus on the $4,923$ onion sites with blockchain addresses. 

\subsubsection{Build a Ground Truth Dataset.}
First, we randomly sample $100$ onion sites, and check the content of all web pages belonging to these domains. In total, we have identified $748$ pages. Then, two of our co-authors manually annotate each web page based on the type of illegal services or goods it provides, following the categories listed in Table~\ref{tab:category}. For pages with only one related service or good, a single label will be assigned. For those related to multiple services or goods, the corresponding multiple labels will be assigned accordingly. For those without illicit activities~(e.g., personal blogs, hosting providers, etc.), the label ``\textit{Other}'' will be assigned. To ensure annotation correctness, we confirm the label only if the two co-authors assign the same label to the same web page. Otherwise, they discuss and further consult a security expert with over 5-year research experience on the dark web to reach a consensus. After all the web pages are labeled, we extract the labels of an onion site by aggregating all the labels of web pages under its domain. Note that if an onion site has multiple labels, we assign the single label ``\textit{Shop}'' for simpler denotation.
In summary, we have identified 55 malicious onions out of the 100 selected ones, and 168 malicious web pages under their domains, as listed in Table~\ref{tab:category}.

\begin{table*}[]
\centering
\caption{The Overview of Ground Truth Dataset.} 
\label{tab:category}
\resizebox{\linewidth}{!}{%
\begin{tabular}{@{}cllcc@{}}
\toprule
Index &
  Category &
  Description &
  \multicolumn{1}{l}{\# Onions} &
  \multicolumn{1}{l}{\# Pages} \\ \midrule
1 &
  Investment Scams &
  \begin{tabular}[c]{@{}l@{}}Cryptocurrency-related investment scams, i.e., Ponzi scams, \\ bitcoin generation scams.\end{tabular} &
  7 &
  9 \\
2 &
  Private Key &
  \begin{tabular}[c]{@{}l@{}}Selling the private keys or partial balance of hacked and \\ abandoned bitcoin wallets.\end{tabular} &
  4 &
  9 \\
3  & Clone Card        & Selling the transfers and clone of prepaid cards. & 12 & 53  \\
4  & Counterfeit Bills & Trading counterfeit bills.                        & 5  & 5   \\
5  & Citizenship       & Selling leaked citizenship databases.             & 1  & 2   \\
6  & Drugs             & Trafficking of prohibited drugs.                  & 1  & 2   \\
7  & Hacker            & Providing hacking services or hacking guides.     & 2  & 3   \\
8  & Hitmen           & Providing murdering services.                     & 1  & 2   \\
9  & Sexual Abuse      & Providing sexual exploitation materials.          & 11 & 15  \\
10 &
  Memberships &
  \begin{tabular}[c]{@{}l@{}}Illicit content communities. Visitors are required to pay \\ membership fees to view the full content.\end{tabular} &
  6 &
  11 \\
11 & Weapons           & Illegal trading of arms and weapons.              & 2  & 2   \\
12 &
  Shop &
  \begin{tabular}[c]{@{}l@{}}Platforms that enable multiple categories of the above \\ illegal trading.\end{tabular} &
  3 &
  55 \\ \midrule
   & Total             &                                                   & 55 & 168 \\ \bottomrule
\end{tabular}%
}
\end{table*}

\subsubsection{Classify Highly Similar Sites.}\label{subsec:cosine_sim}
Our initial exploration suggests that many illicit sites share similar structures and contents. Thus, based on our labeled ground truth dataset, we intend to identify illicit sites that are highly similar to them. 
To measure the similarity between two pages, we adopt the cosine similarity which is commonly used in measuring the text similarity between documents, such as articles and HTML files. Given an onion site, we calculate the cosine similarity between it and all the $12$ classes, and assign its label to the class with the highest score. 
To calculate the cosine similarity between a pair of pages, we construct a feature vector pair ($V_1$, $V_2$). More specifically, we first remove the tag and parse the contents from the HTML files of a page into a word sequence $S$.  Pages from non-English sites are first translated into English by Google Translation API before being converted into word sequences. To further exclude irrelevant textual information, we remove punctuation, numbers, stop words, and words shorter than $3$ characters. 

After preprocessing, we derive a word sequence pair ($S_1$, $S_2$), from which the analyzer further constructs the vector pair ($V_1$, $V_2$). The vector dimensions are equal to the number of distinct words that appear in the two sequences, and each dimension field denotes the frequency of a word in the sequence. Then the cosine similarity is computed from the following equation: 

\begin{equation}
\cos\langle V_1, V_2 \rangle=\frac{\sum_{i=1}^n v_{1_i}v_{2_i}}{\vert{V_1}\vert\cdot\vert{V_2}\vert}
\end{equation}

We set the default labels of the onions sites as ``\textit{Other}''~(i.e., unlabeled), and re-assign a label if they are similar to a certain class. Precisely, we utilize a practically-identified threshold to measure the similarity between a certain class and the target sample. To derive the optimal threshold value, we utilize a randomly selected set of 20 unlabeled onion sites, and apply the classification algorithm with various threshold values. We identify that the threshold value of 0.5 achieves the best classification performance. If an onion site shows similarity with multiple classes (i.e., the similarity scores exceed the threshold for multiple classes), the label from the class with the highest score will be assigned to it. After this labeling process, $2,393$ onion sites with $8,759$ pages are re-assigned a label, while $2,475$ sites remain unlabeled. 

\subsubsection{TF-IDF Based Classification.}
To further complement the cosine similarity, we also utilize the TF-IDF (Term Frequency–Inverse Document Frequency) method to classify onion pages. 
This method is capable of capturing more illicit sites that are content-wise distinct (i.e., lower cosine similarity) but topic-wise intrinsically similar~(i.e., similar keywords and themes). 

TF-IDF measures the uniqueness of a word to a document against a corpus. To classify the onion sites based on TF-IDF, we need to generate a feature set containing keywords to represent the characteristics of each class. 
We first generate a word sequence for each labeled page, using the same approach as detailed in Section~\ref{subsec:cosine_sim}. Next, we concatenate the word sequences from the same class into one document, generating 12-word sequence documents respectively. Then, we feed them to a TF-IDF Vectorizer, which will calculate a TF-IDF vector as the representation for each document. In this vector, each dimension field is a TD-IDF value denoting the weight of a word in the document. To reflect the most relevant features in each class, we extract $20$ keywords with the highest weight in each document, and merge them to form a 220-word feature set. As an illustration, we list the $10$ most relevant words in Table~\ref{tab:keyword}. 

\begin{table*}[t]
\centering
\small
\caption{Top 10 Keywords of Each Category.} 
\label{tab:keyword}
\resizebox{\linewidth}{!}{%
\begin{tabular}{@{}cl@{}}
\toprule
Category          & Keywords                                                                                                                              \\ \midrule
Investment Scams &
  \begin{tabular}[c]{@{}l@{}}flaw, multiply, bitcoins, client, transaction, innovative, digital, investment, \\ found, history\end{tabular} \\
Private Keys       & balance, electrum, privkey, lordpay, wallet, private, wallets, key, accounts, price                                                   \\
Cloned Card        & buyed, says, transfer, product, cards, money, western, card, paypal, union                                                            \\
Counterfeit Bills & bills, euro, value, price, usd, bill, dollar, amounts, costs, ship                                                                    \\
Citizenship &
  \begin{tabular}[c]{@{}l@{}}dateofbirth, addresscity, ahmet, erdogan, motherfirst, addressdistrict, \\ addressneighborhood, birthcity, doororentrancenumber, fatherfirst\end{tabular} \\
Drugs             & name, courses, middle, effects, extreme, zip, contacts, city, country, drug                                                           \\
Hacking            & \begin{tabular}[c]{@{}l@{}}hack, tutorials, programs, confirmation, begin, archive, automatically, \\ message, zip, send\end{tabular} \\
Hitmen           & \begin{tabular}[c]{@{}l@{}}target, hitmen, provide, information, identifying, identify, murder, profile, \\ address, job\end{tabular} \\
Sexual Abuses      & porno, porn, video, sex, free, teen, film, gay, online, russian                                                                       \\
Memberships       & \begin{tabular}[c]{@{}l@{}}access, pin, payment, redirect, page, deposit, authorization, creation, \\ reversed, red\end{tabular}      \\
Weapons           & \begin{tabular}[c]{@{}l@{}}darkseid, armour, calibers, drkseid, modify, succesfully, arms, guns, years, \\ expertise\end{tabular}     \\
Shops              & onion, index, hidden, marketplace, cards, tor, card, credit, hosting, service \\ \bottomrule
\end{tabular}%
}
\end{table*}

For the remaining unlabeled onion sites, we utilize the constructed feature set to further filter out those intrinsically similar to the labeled ones. We first use the same approach to construct the TF-IDF vectors for the unlabeled sites. 
Next, we project the TF-IDF vectors onto the feature set to generate new feature vectors for this task. Then we calculate the cosine similarity for the labeling. After labeling, 116 onion sites with 847 pages are reassigned as a class label. 

\subsubsection{Classification Results.} 
\label{subsec:classification_results} 
As is shown in Table~\ref{tab:classification}, through our three-phase classification approach, we have identified $2,564$ onion sites distributed across 12 types of illegal activities, out of the total $18,014$ onion sites. 
Considering the stringent similarity comparison criteria adopted in our approach for classification, we have observed low false positive/negative rates. 
As further confirmation, we assemble a small dataset containing 100 randomly selected unlabeled sites. We manually check their contents and verify that 96 sites of them are indeed unlabeled as predicted. We will include a more detailed discussion in Section~\ref{sec:limit}. 

\begin{table*}
\centering
\small
\caption{Classification of Onion Sites and Addresses.}
\label{tab:classification}
\begin{tabular}{@{}ccccc@{}}
\toprule
Category &
  \# Onions &
  \# Pages &
  \begin{tabular}[c]{@{}c@{}}\# BTC \\ Addr.\end{tabular} &
  \begin{tabular}[c]{@{}c@{}}\# BTC addr.\\ (Illicit)\end{tabular} \\
  \midrule
Clone Card        & 447  & 2,493 & 362  & 362     \\
Investment Scams  & 1,254 & 1,498 & 6,851 & 231    \\
Sexual Abuses      & 392  & 1,147 & 334  & 334   \\
Memberships       & 133  & 242  & 103  & 100  \\
Hacker            & 55   & 69   & 101  & 45         \\
Shop              & 94   & 4,048 & 68   & 54      \\
Counterfeit Bills & 66   & 71   & 15   & 14   \\
  Weapons          & 6    & 11   & 4   & 5\\
 Drugs            & 18   & 45   & 5   & 5  \\
  Private Key      & 90   & 138  & 89  & 10\\ 
 Citizenship      & 3    & 4    & 1   & 1 \\
 Hitmen          & 6    & 8    & 5   & 4 \\
 Multi-Categories & -    & -    & 200 & 24 \\
 Total            & 2,564 & 9,774 & 304 & 1,189\\
  \bottomrule
\end{tabular} 
\end{table*}

\subsection{Identifying Illicit Blockchain Addresses}
\label{sec:illicit-addr}
Among the identified $8,138$ Bitcoin addresses and $1$ Ethereum address contained in the $2,564$ illicit onion sites, we retain those related to~(i.e., linked to or owned by) the onion owners. For this purpose, we filter out and remove the irrelevant addresses based on the following criteria: 1) leaked or hacked addresses sold by the sites (in the Private Key category); 2) listed addresses that have transactions before with the site's payment address, which usually appears in investment scam sites; 3) addresses appeared on the forums or message boards of the site, which is typically owned by the visitor rather than the site; and 4) other unrelated but address-formatted hashes. After the filtering, we retain $1,189$ Bitcoin addresses owned by illicit onion sites. No illicit Ethereum address remains in this step. 

\subsection{Extracting Blockchain Transactions}
To calculate the illicit income, we obtain and investigate the transaction records of illicit addresses. The transactions of blockchain addresses are publicly available, owing to the transparency nature of decentralized ledgers. We use the transaction data API provided by BTC.com~\cite{btccom} to achieve all history transactions of the $1,189$ illicit Bitcoin addresses. Among them, $784$ addresses have no transaction records, indicating that they have never received any BTC. Other $405$ addresses have incurred $16,254$ transactions in total. 

\begin{table*}[t]
\centering
\small
\caption{The Statistics of Found URLs.}
\label{tab:surface-website}
\begin{tabular}{@{}ccc|ccc@{}}
\toprule
Category          & \# URLs & \# Addresses & Category & \# URLs & \# Addresses \\ \midrule
Clone Card        & 4       & 4            & Hacker   & 1       & 1            \\
Investment Scams  & 6       & 5            & Shop     & 1       & 1            \\
Sexual Abuses            & 30      & 1            & Forums   & 1       & 1            \\
Counterfeit Bills & 1       & 1            & Total    & 44       & 14           \\ \bottomrule
\end{tabular}%
\end{table*}

\begin{figure}[t]
  \centering
  \includegraphics[width=\linewidth]{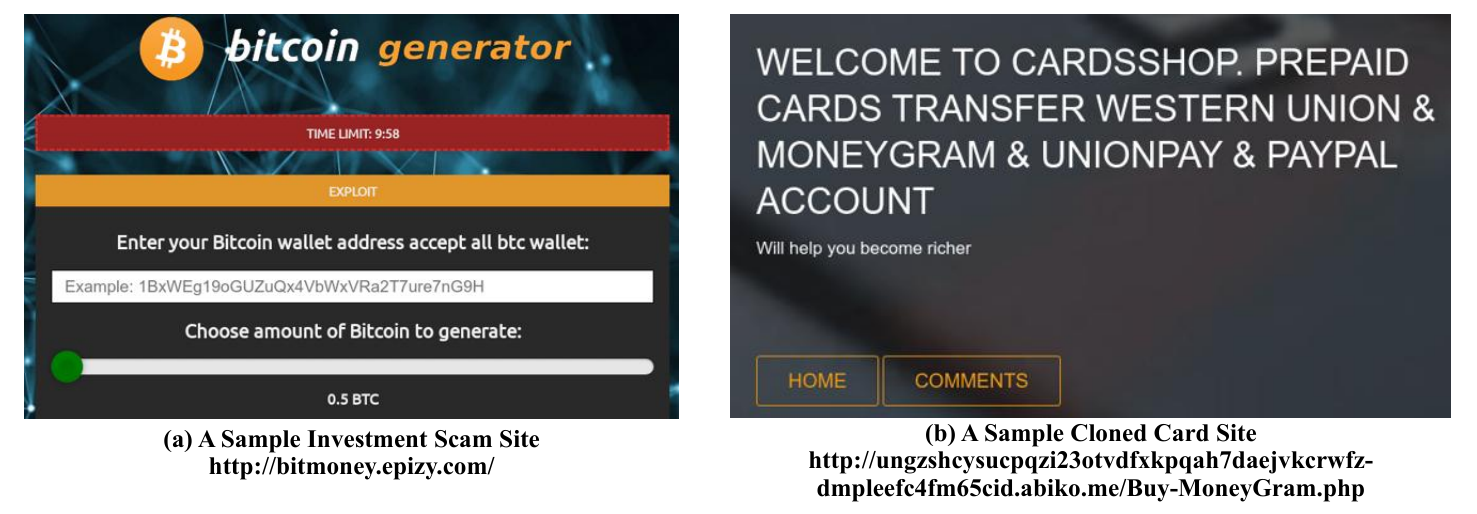}
  \caption{The Screenshots of Two Sites on the Surface Web.}
  \label{fig:scam-screenshot}
\end{figure}

\subsection{Tracing Illicit Blockchain Addresses on the Surface Web}
As the illicit blockchain addresses may participate in illicit activities that can be exposed on the surface web, we also search them using Google Search API to trace their activities. During the searching process, we filter out blockchain explorer results since they only contain the addresses' on-chain activities that are already obtained in the previous step. After this searching process, we obtain another $111$ distinct URLs from $42$ addresses. 

By manually inspecting the web contents of the $111$ URLs, we identify $47$ abuse reports. These abuse reports are scattered across various sources, including news, blogs, and social media posts, Bitcoin abuse databases~(we consider major popular databases, such as BitcoinAbuse~\cite{bitcoinabuse}, BitcoinWhosWho~\cite{bitcoinwhoswho}, Cryptscam~\cite{cryptscam}, and hashXP~\cite{hashxp}), and even court files. 

For the remaining $64$ URLs, we find that $44$ of them are related to illicit activities, as listed in Table~\ref{tab:surface-website}. 
We provide two screenshots of found sites in Figure~\ref{fig:scam-screenshot}, which are related to investment scams and cloned card illicit activities. Among the $44$ URLs, $26$ of them come from blogs or file-sharing services, exploited by perpetrators to spread illicit content. We thus consider these service operators to be irrelevant to the illicit addresses. The rest $18$ illicit URLs can be traced to the website operators' identities, including information on their IP addresses, domain registrant names, etc. In more detail, we utilize Urlscan.io~\cite{urlscan} and Whois database to derive the identity information from $18$ URLs. 

\section{Evaluation} 

In this section, we evaluate the impacts of cryptocurrency abuses and the criminal campaigns behind the scenes on the dark web ecosystem. Our evaluation is driven by the following research questions (RQs):

    \noindent\textbf{RQ1. What types of illicit activities exist on the dark web, and what impacts do they have on the dark web ecosystem?} Existing studies and media reports have analyzed specific illicit activities on the dark web. There are also open-source abuse databases that include a small number of illicit blockchain addresses related to the dark web. However, there is a lack of an in-depth understanding of illicit activities and cryptocurrency abuses on the dark web.
    
    To answer RQ1, we apply the proposed approach~(Step 2-6 detailed in Section~\ref{sec:study-design}) to systematically identify the illicit sites and blockchain addresses from our dataset. We examine the landscape of cryptocurrency-related abuses on the dark web based on their high-level characteristics, as detailed in Section~\ref{sec:overview}.
    
    \noindent\textbf{RQ2. Who are behind the illicit onion sites?} Simply characterizing the illicit activities from the onion sites or blockchain addresses individually is insufficient to understand the perpetrators, as they can remain hidden just like the main body of an iceberg remains underwater. Clustering can further interconnect the individual sites and addresses, which provides a broader and clearer view of the illicit campaigns launched by those perpetrators. However, the existing web page clustering techniques are not directly applicable to onion pages. On the one hand, the prevalent page content duplication brings uncertainty to content-based approaches. On the other hand, the identity of dark web perpetrators is protected by anonymous networks, making it impossible to extract features~(e.g., domains and IP addresses) for clustering.
    
     To answer RQ2, we propose an approach to cluster illicit onion sites based on identifiers such as blockchain addresses, email addresses, and identity information, as presented in Section~\ref{sec:cluster}. Considering the identifiers shared among the onion sites inside each cluster, we thus consider a cluster as an illicit campaign hereafter.
    
    \noindent\textbf{RQ3. What are the characteristics of illicit campaigns on the dark web?} Based on the illicit campaigns identified in RQ2, we strive to answer the following sub-questions to further characterize them: What illicit goods or services do they offer on their onion sites, and is it a single category or multiple categories? Which goods or services are more popular and profitable on the dark web? What methods do they use to increase their illicit income? Do they run several sites across the dark and surface web to conduct illicit activities?
    
    To answer RQ3, we conduct a multi-dimensional analysis to characterize the illicit campaigns, as presented in Section~\ref{sec:campaign}.

\section{RQ1. Illicit Activities and Impacts on the Dark Web}
\label{sec:overview} 
In this section, we characterize the illicit activities based on the dataset constructed in Section~\ref{sec:study-design}, which contains 1,189 Bitcoin addresses and 1,246 email addresses from 2,564 identified illicit onion sites spanning 12 types of illicit activities.  
First, we evaluate the illicit income received by the blockchain addresses. Next, we adopt a multi-dimensional analysis to study the characteristics of illicit onion sites, blockchain addresses, and email addresses separately. For simplicity, we refer to blockchain address as \textit{address} hereafter, unless stated otherwise.

\begin{figure}[ht]
\centering 
\subfigure[The Income of Each Category.]{
    \label{fig:btc-type}
    \includegraphics[width=0.38\linewidth]{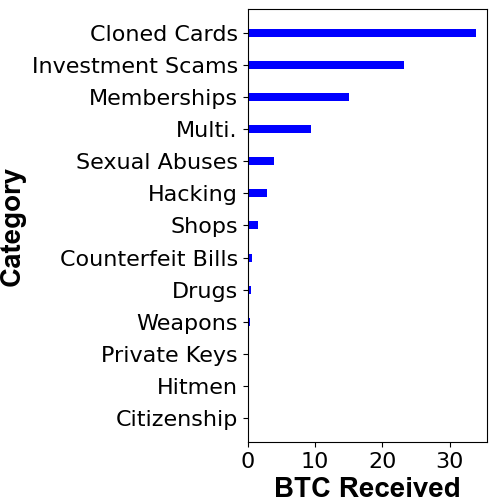}
}
\subfigure[The Income of Each Address.]{
    \label{fig:tx-btc-dis}
    \includegraphics[width=0.27\linewidth]{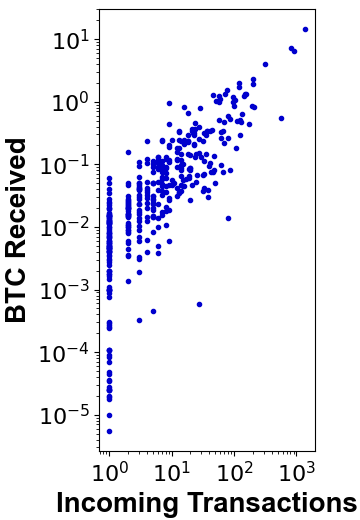}
}
\subfigure[The Distribution of Addresses' Active Periods.]{
    \label{fig:active-day}
    \includegraphics[width=0.27\linewidth]{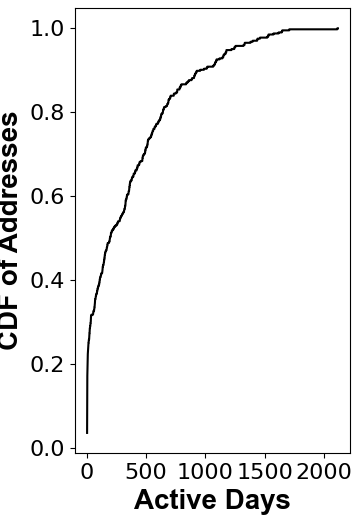}
}
\caption{The Income and Active Periods of Illicit Addresses.}
\label{fig:income-active-period}
\end{figure}

\subsection{Illicit Income}
We estimate the illicit income from $16,254$ distinct transaction records obtained from the $1,189$ illicit addresses. Among them, $405$ addresses have received $90.81$ BTC in $10,996$ incoming transactions by study time. $351$ addresses ($86.67\%$ of the $405$ addresses) have transferred all their illicit income away, while only $1.17$ Bitcoins remain in the rest of the addresses. When estimating illicit income, we exclude $284$ internal transactions (transactions that occur between illicit addresses) to avoid duplicate values. We note that the estimated income volume is lower than that reported in \cite{lee2019cybercriminal} and \cite{chainanalysis}, plausibly due to the slight difference in our focused scope. We target the ongoing campaigns of cryptocurrency-related illicit activities, which is indeed the lower bound and subset of the total illicit activities on the dark web. In addition, the estimated illicit income is from our confirmed malicious addresses, without expanding to potentially malicious or suspicious addresses as mentioned in the two works. 
We estimate the BTC value flowing into each illicit activity to evaluate the financial impacts of different illicit activities. Figure~\ref{fig:btc-type} displays the Bitcoin received by illicit addresses from each category. Cloned card trading sites make the most profit and receive $33.84$ BTC. Citizenship services receive the least, where only one address receives a small transfer. 

\paragraph{Volumetric Features} To understand the volumetric features of illicit transactions, we present the distribution of incoming transactions and values received by each address in Figure~\ref{fig:tx-btc-dis}. One of the 405 addresses receives only one internal transfer, and it is thus excluded. We observe from the figure that most of the addresses only receive a small number of transactions (less than 100). The received amount is mainly below 1 BTC. 

\paragraph{Temporal Features} After inspecting the BTC received and sent, we find that the illicit addresses have been more active~(i.e., with a larger number and amount of transactions) since 2018. One plausible reason for this observation is that many onion sites created earlier have been deactivated by the time of our study.

\subsection{Active Periods} 
We use the days between the earliest and latest transaction time of an address to represent its active period, as shown in Figure~\ref{fig:active-day}. $62.56\%$ of the addresses have an active period of less than one year. $41$ (about $10\%$) of the addresses are active for only one day, as they receive only a few transfers within a day and immediately transfer out the income. 
Since the initial transaction time for a few addresses (i.e., as early as September 2015) predates the start of the scraping, their active periods are longer than the duration of our onion site collection, as shown in Figure~\ref{fig:active-day}. By correlating the active period with the illicit income of the addresses, we identified a highly imbalanced distribution pattern. More specifically, the 30 long-living addresses~(active for more than three years) are responsible for more than 56\% of the total illicit revenue gathered in 5372 transactions, while those active for shorter periods~(less than one year) correspond to about 13\% of the revenue. Regarding the transaction numbers, the long-living addresses incur an average of 21\% more per transaction than the short-term ones. 

Note that we exclude the dormant addresses, as they never empty their illicit income given the minimal cumulative amount received (i.e., the median value of 0.0018 BTC per address). More specifically, we exclude $21$ addresses, accounting for $0.29$ BTC received in 48 transactions. We manually verify that these addresses are used to receive payments on the illicit onion sites, confirming they are indeed illicit. 

\subsection{The Most Profitable Addresses}
\label{sec:profiable-addr}
The top five profitable addresses account for over one-third of the total received illicit income, as listed in Table~\ref{tab:profitable-addr}. We provide a brief discussion on the top three addresses. The first address has received $14.64$ BTC in $1,364$ transactions from October 4, 2015. It is the payment address of $16$ identified investment scam sites. The second one is the payment address of an onion site offering child pornography content to its members, and it requires its customers to pay monthly membership fees for full access. The third one is a payment address across three categories of illicit sites. More details will be discussed in Section~\ref{subsec:abuse_rep}. 

\begin{table*}[]
\centering
\small
\caption{Top 5 Profitable Addresses.}
\label{tab:profitable-addr}
\resizebox{\linewidth}{!}{%
\begin{tabular}{@{}lccc@{}}
\toprule
Address &
  Category &
  \begin{tabular}[c]{@{}l@{}}\# Incoming\\ Transactions\end{tabular} &
  \begin{tabular}[c]{@{}l@{}}BTC\\ Received\end{tabular} \\ \midrule
1CHvWk36MR5aCz72jViS7jSub9utJf3jii & Investment Scams & 1,364 & 14.64 \\
1EKrfiWZoABz17DWJxUrycQKg3Fo4zZ2Z2 & Memberships      & 808   & 7.29  \\
1Gs7Aztizk2rNNSE6AbpK4K7yAFTCZKV9a &
  \begin{tabular}[c]{@{}l@{}}Cloned Card, \\ Sexual Abuses, \\ Memberships\end{tabular} &
  746 &
  5.84 \\
1ENrJ77ubXo5eeip2XpohC4jQgKwLWxfuA & Cloned Card       & 242   & 3.81  \\
1FvftoUHVpFZaasCnPTUcbccqi5PYnm5S6 & Cloned Card       & 180   & 2.22  \\ \bottomrule
\end{tabular}%
}
\end{table*}

\subsection{Multi-category Addresses}
We have identified $23$ addresses as the payment addresses for onion sites across multiple categories. Most of the multi-category addresses involve two to three types of illicit activities. One interesting observation is that nearly all of them (only one exception) target cloned card trading or investment scams which are the two most profitable categories. We will further analyze the connections between these two categories in Section~\ref{sec:speculator}. The only exception targets memberships and sex abuse sites instead.

Among the $23$ multi-category addresses, the address `1KRkAW\-DH5q7U5rdTM1rREmepk1pxpCAVKE' is the most prominent example. This address is found in $69$ illicit sites across eight categories (i.e., Counterfeit Bills, Hacking, Sexual Abuses, Memberships, Weapons, and Shops). It is unusual for an individual seller or dark web marketplace to create diversified pages selling different types of goods. Instead, such sites are scam sites with a high probability. We detail scam reports of the payment addresses in Section~\ref{subsec:abuse_rep}. 

\subsection{Multi-payment Address Scenarios}
We find $319$ onion sites in our dataset configured with multiple payment addresses. They can be further split into two scenarios. The first is to set different addresses for different items. For example, a membership site named ``redroo6dogiqe\-gxn.onion'' provides live streams and videos of murder, rape, and torture. It offers several access levels at different prices. For each level, it sets a specific payment address. The second is that the site randomly picks one of the owned addresses when displaying the payment address for an order. For example, a site named ``vsmgkgnltwzrc7tx.onion'' sells hacking guides and owns seven payment addresses. When displaying a payment address QR code, its page runs a script to randomly pick one from all of its addresses.

\subsection{Vanity Onion Names}
An onion name typically looks random--it is generated by hashing a public key. However, the creator can generate an onion name with human-readable parts using brute-force methods, which is called a vanity onion name~\cite{vanity_name}. Onion addresses with the same distinguishable vanity name suggest that they may be created by the same entity, which helps us detect and link potential illicit campaigns on the dark web.
We have observed illicit onion sites with distinguishable vanity names showing prefixes that are usually identified as the shop names, the goods or services offered, and their variants. For instance, $66$ onion domain names, which are linked to an escrow marketplace called DeepMarket, start with ''deepmar"~(e.g., deepmar27rpxago5.onion, deepmar3k3qtzszd.onion). After a closer look at these onion site contents, we further discover that all of them embed the bitcoin address ``1AYnoFpTbfVXYpADgzidDCJHE1X5UhyGqu''. 

\subsection{Email Services}
The onion service itself will not capture and use the operator’s private information, but displaying an email address from a non-anonymous email service can expose the operator's true identity. We examine the email services used by illegal website operators, and find that most illicit operators are cautious to choose an anonymous email service. The two most used email services are \texttt{SecMail} and \texttt{ProtonMail}, with $299$ and $269$ email addresses found respectively. They are both popular encrypted email service providers. In addition, we found $162$ email addresses with the domain name ``email4tor.com''. We believe that this is an email service specially targeting dark websites.

\vspace{0.1in}
\noindent\fbox{
	\parbox{0.95\linewidth}{
		\textbf{Answer to RQ1:} 
		Based on the blockchain transactions, we detect transactions amounting to 90.8 BTC. Among the 12 types of illicit activities, cloned card sales are the most prevalent, accounting for 33.4 BTC from 341 sites. Despite the common cryptocurrency abuses in the dark web, 88.6\% of illicit addresses are inactive or receive only a few~(i.e., 10 or fewer) transfers.
}}
\vspace{0.1in}

\section{RQ2. Identifying the Campaigns Behind Illicit Onions}
\label{sec:cluster}

In this section, we aim to reveal the campaigns behind illicit onion sites according to their underlying connections. We first detail the clustering approach to identify the illicit campaigns, based on several features extracted from onion sites including bitcoin addresses, email addresses, and identity information. Furthermore, we examine and characterize the identified campaigns. 

\subsection{Clustering Approach}
The clustering approach consists of three concatenated phases. In each phase, the onion sites are clustered based on one of the following features specifically, i.e., the Bitcoin address, email address, and domain name information retrieved from the surface web. The results of each phase will be passed to the next phase for further clustering.
Note that we do not expand and cluster new addresses based on existing illicit addresses using multi-input and change heuristics, compared to that mentioned in \cite{lee2019cybercriminal}. Since the two techniques could introduce false positives for clustering as pointed out by~\cite{goldfeder2017cookie,moser2021resurrecting}, we thus mainly focus on the underlying connections among confirmed illicit addresses.
\subsubsection{Bitcoin Address Clustering.} 
We cluster onion sites based on their Bitcoin addresses first. The links revealed in the addresses' shared onion sites and transaction records are taken into account in this phase. Therefore, the clustering is split into three consecutive steps, as described below:

\paragraph{Shared Onion Sites} Two Bitcoin addresses are clustered into a group if they are both the payment addresses of an illicit onion site. Based on $192$ such addresses, we generate $77$ clusters from $1,189$ onion sites. The remaining $1,375$ sites stay isolated, indicating that they do not have shared payment addresses with each other.

\paragraph{Common Inputs} Inspired by the multi-input heuristic~\cite{meiklejohn2013fistful}, we cluster Bitcoin addresses together if they are the common inputs of a transaction. The heuristic assumes that input addresses in a transaction belong to the same entity, which is commonly used in address clustering. We find common inputs in $523$ transactions, and cluster $56$ additional onion sites into groups based on them. In this phase, the $77$ clusters are further merged into $67$ clusters, and $12$ new clusters are generated. $1,319$ illicit onion sites stay isolated. Before proceeding to the next step, we also scan for the transactions related to mixing services, in particular JoinMarket transactions, to eliminate the irrelevant input addresses in one transaction. Utilizing the algorithm proposed by Goldfeder et al.\cite{goldfeder2017cookie}, we find no such instance among the $523$ transactions.
\paragraph{Internal Transactions} If two Bitcoin addresses are the input and output of an internal transaction, then they are clustered together. Internal transactions may arise from addresses that gather and transfer illicit funds, revealing the correlations among illicit addresses. We find $22$ addresses in $85$ internal transactions. They group four additional onion sites into clusters. In this step, $79$ clusters are merged into $72$ clusters and no new clusters are generated. $1,315$ illicit onion sites stay isolated.
We have identified correlations among the existing clusters, despite a stable number of isolated sites in this step. 
\subsubsection{Email Address Clustering.}
In this phase, we cluster the onion sites based on the connections related to their email addresses, as many onion sites without a blockchain address on the web page may otherwise provide an email address. Furthermore, these sites may share email addresses with ones containing blockchain addresses, suggesting that they may be under the control of the same entity or group. Clusters that do not contain any blockchain addresses are excluded from the results, as we are unable to further estimate the illicit incomes for them due to the absence of the addresses. From the email clustering, we cluster additional $146$ onion sites according to $46$ email addresses. Note that $137$ out of the $146$ previously unlabeled onion sites do not contain any blockchain address.
\subsubsection{Identity Information Clustering.} 
The domain names, including IPs, registrants, etc., carry more information related to the controlling entities behind a set of colluding websites. In this phase, we aim to identify such websites containing the same IP or registrant name. 
From them, we group all the blockchain addresses included in those websites into a single cluster.  Furthermore, we aggregate the onion sites containing any of these blockchain addresses into the same cluster. Note that, during grouping, we find some IPs or registrant names related to a large number of domains. We manually inspect them to check whether they belong to public services and exclude the public ones. 
This step merges two of the previous clusters into one, with one extra onion site and its Bitcoin address included. We obtain a total of $66$ clusters after this phase. 

\subsection{Results}
\begin{figure}[h]
  \centering
  \includegraphics[width=0.95\linewidth]{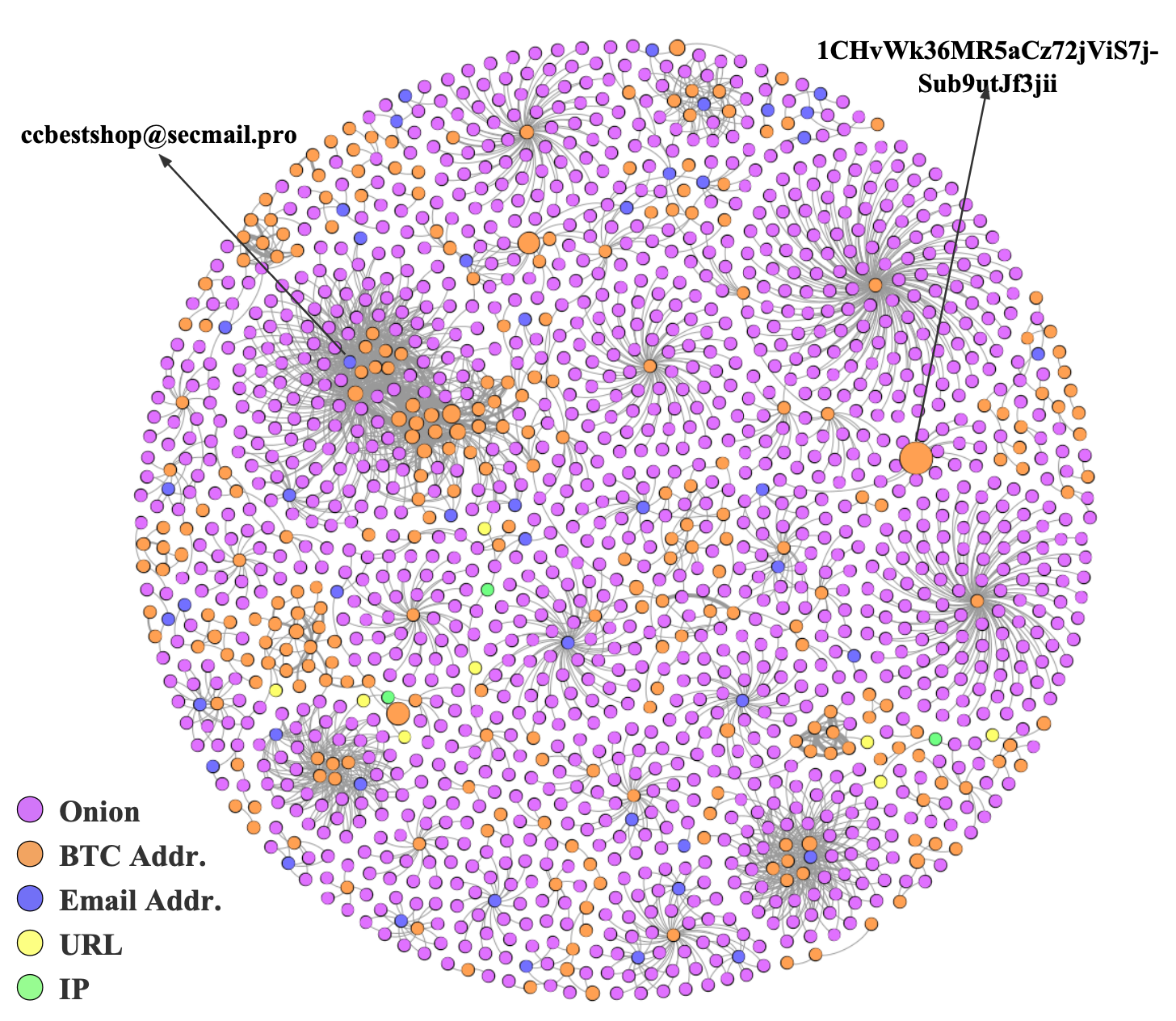}
  \caption{The Clusters of Onion Sites.} 
  \label{fig:cluster-graph}
  \vspace{-0.3cm}
\end{figure}

The result of each clustering phase is summarized in Table~\ref{tab:clustering}. After this process, we derive $66$ clusters from $1,396$ onion sites. The clustered addresses have received $80.72$ BTC, which accounts for $88.9\%$ of the total illicit income. 
To visualize the clustered onion sites and the related addresses~(i.e., Bitcoin addresses, email addresses, URLs, and IPs), we plot this information as nodes in Figure~\ref{fig:cluster-graph}. Each colored node denotes a type of address, and the edges between them indicate their association through clustering. Specifically, the Bitcoin address node sizes are proportional to the transaction amount. In total, we have identified $1,725$ nodes in the 66 clusters. 

\begin{table*}[t]
\caption{The Clustering Results. }
\label{tab:clustering}
\resizebox{\linewidth}{!} {
\begin{tabular}{@{}cccccc@{}}
\toprule
Phase                    & \# Clusters & \# Onions & \# Bitcoin Addr. & \# Email Addr. & \# IPs \\ \midrule
Bitcoin Address Clustering-I     & 77            & 1,189     & 192          & -         & -      \\
Bitcoin Address Clustering-II    & 79            & 1,245     & 258          & -         & -      \\
Bitcoin Address Clustering-III   & 72            & 1,249     & 262          & -         & -      \\
Email Address Clustering         & 67            & 1,395     & 271          & 46        & -      \\
IP Clustering & 66            & 1,396     & 272          & 46        & 3      \\ \bottomrule
\end{tabular}%
}
\end{table*}

\subsubsection{Cluster Size}

We use the number of onion sites and Bitcoin addresses included in a cluster to represent its size. More than half (36) of the clusters contain only one Bitcoin address, indicating the campaigns behind them tend to create multiple onion sites for illicit purposes with only one address to receive payments. 26~(39.4\%) of the clusters operate 10 or more illicit onion sites, which is a plausible strategy to increase their exposure on the dark web.

\subsubsection{Most Profitable Clusters}
\label{profitable-cluster}
 The top 10 most profitable clusters are listed in Table~\ref{tab:profitable-family}. The first cluster also contains the largest number of sites and blockchain addresses, accounting for $26.90$ BTC received from $2,099$ transactions. For easy reference, we mark one of its email addresses~(i.e., ccbestshop@secmail.pro) in Figure~\ref{fig:cluster-graph}. This email address appears as a contact for 92 sites. We will present more details of the cluster in Section~\ref{sec:campaign}. The second cluster consists of a single bitcoin address that appears on 31 investment scam sites, as marked in Figure~\ref{fig:cluster-graph}. This address is also the most profitable, as given by Table~\ref{tab:profitable-addr}.

\begin{table*}[t]
\centering
\caption{Top 10 Profitable Campaigns. }
\label{tab:profitable-family}
\resizebox{\textwidth}{!}{%
\begin{tabular}{@{}cccccccc@{}}
\toprule
No. &
  Onion Sites &
  \# &
  Category &
  \# Bitcoin Addr. &
  \# Email Addr. &
  \# URLs &
  BTC Received \\ \midrule
1 &
  \begin{tabular}[l]{@{}c@{}}bqt2karhqo3wlubd.onion, \\ c3icqzp7yla3z2ap.onion...\end{tabular} &
  218 &
  \begin{tabular}[c]{@{}c@{}}Memberships, Cloned Card,\\ Drugs, Investment Scams\end{tabular} &
  49 &
  7 &
  4 &
  27.26  \\
2 &
  \begin{tabular}[c]{@{}c@{}}22222222sty2trl2.onion, \\ 22222222gkxknocu.onion...\end{tabular} &
  31 &
  Investment Scams &
  1 &
  1 &
  0 &
  14.64 \\
3 &
  \begin{tabular}[c]{@{}c@{}}cpcpcp5mohuba7in.onion, \\ cardssbgdppyt4pw.onion...\end{tabular} &
  21 &
  \begin{tabular}[c]{@{}c@{}}Memberships, Hitmen,\\ Cloned Card, Sexual Abuses,\\ Investment Scams\end{tabular} &
  26 &
  0 &
  1 &
  11.31 \\
4 &
  \begin{tabular}[c]{@{}c@{}}ylffvocck35ov4tu.onion, \\ teenpuvpovgmpygq.onion...\end{tabular} &
  56 &
  \begin{tabular}[c]{@{}c@{}}Sexual Abuses, Memberships,\\ Cloned Card, Investment Scams\end{tabular} &
  18 &
  3 &
  0 &
  7.74  \\
5 &
  \begin{tabular}[c]{@{}c@{}}be7gb2cmlr5vlrsy.onion, \\ pedonetay6emfuww.onion\end{tabular} &
  2 &
  Memberships &
  1 &
  0 &
  0 &
  1.92 \\
6 &
  \begin{tabular}[c]{@{}c@{}}o3fkhst6gymb6duo.onion, \\ ub2fkkeeyl3jgwwz.onion\end{tabular} &
  2 &
  Cloned Card &
  2 &
  0 &
  0 &
  1.91 \\
7 &
  \begin{tabular}[c]{@{}c@{}}5snp5ngqa75lzf2i.onion, \\ nrclqxgvvait5cv2.onion...\end{tabular} &
  10 &
  \begin{tabular}[c]{@{}c@{}}Weapons, Cloned Card,\\ Investment Scams, Hacking\end{tabular} &
  13 &
  1 &
  0 &
  1.88 \\
8 &
  \begin{tabular}[c]{@{}c@{}}btcgenmtbimjhlco.onion, \\ bitcoinrmnuijyli.onion...\end{tabular} &
  5 &
  \begin{tabular}[c]{@{}c@{}}Private Key,\\ Investment Scams\end{tabular} &
  3 &
  1 &
  0 &
  1.45 \\
9 &
  \begin{tabular}[c]{@{}c@{}}countrfoioed4ckx.onion, \\ cardsm4fgcw5po3v.onion...\end{tabular} &
  5 &
  \begin{tabular}[c]{@{}c@{}}Hacker, Clone Card,\\ Counterfeit Bills\end{tabular} &
  5 &
  0 &
  0 &
  1.38  \\
10 &
  \begin{tabular}[c]{@{}c@{}}gieg73fs3polsjwx.onion, \\ za7z6ht74hcndhxj.onion...\end{tabular} &
  40 &
  Cloned Card &
  10 &
  2 &
  1 &
  0.99 \\ \bottomrule
\end{tabular}%
}
\end{table*}

\vspace{0.1in}
\noindent\fbox{
	\parbox{0.95\linewidth}{
		\textbf{Answer to RQ2:} 
		 We have identified 66 campaigns behind the 1,396 illicit onions in our dataset through the clustering analysis. They hold 272 Bitcoin addresses and receive 80.7 BTC, accounting for 88.9\% of the total illicit income. 54.6\% of the campaigns utilize one address to receive incoming payments. To gain more exposure on the dark web, 39.4\% of campaigns operate 10 or more illicit sites.
}}
\vspace{0.1in}

\section{RQ3. Characterizing Illicit Campaigns on the Dark Web}
\label{sec:campaign}
In this section, we present a multi-dimensional characterization of the illicit campaign on the dark web, from the perspectives of their illicit activities on the dark web~(Section~\ref{sec:speculator}), their correlations with the surface web~(Section~\ref{sec:surfaceweb-activity}), and features identified from abuse reports~(Section~\ref{subsec:abuse_rep}). 

\subsection{Activities on Speculator-oriented Markets}
\label{sec:speculator} 
We list the number of campaigns involved in each category of the 12 illicit activities, as shown in Table~\ref{tab:campaign-activity}. Note that one campaign may participate in multiple illicit activities, just like many legitimate businesses attempt by offering a range of items or services for sale. 
Most campaigns~(46 out of 66) target cloned card sales and investment scams due to their high profitability. They commonly leverage irresistible deals to lure the victims into their traps.  
Specifically, We identify a large-scale campaign mainly targeting cloned card sales, which generates the highest illicit revenue among all campaigns. It consists of 42 Bitcoin addresses and 193 sites (191 onion sites and 2 normal sites) that are dedicated to selling prepaid cards, Western Union money transfers and PayPal accounts at a much lower price. This campaign is also involved in other 3 types of illicit activities, Investment Scams, Drugs, and Memberships. Moreover, we further reveal its business network operated on the surface network, to be detailed in Section~\ref{sec:surfaceweb-activity}.

In addition, it is uncommon for the campaigns to target various illicit activities. For example, we find that 14 out of the 26 campaigns targeting investment scams also operate onion sites in other categories, ranging from counterfeit bill sales to child pornography services. There are two plausible explanations for the finding: 1) the campaigns try to generate income via scams; 2) they indeed offer a wider range of illicit goods and services to harvest higher illicit revenue. 

\begin{table}
\centering
\small
\caption{Number of Campaigns for Illicit Activities.}
\label{tab:campaign-activity}
\begin{tabular}{@{}cc|cc@{}}
\toprule
Category          & \# Campaigns & Category    & \# Campaigns \\ \midrule
Cloned Card        & 26           & Shops        & 5            \\
Investment Scams  & 26           & Drugs       & 2            \\
Memberships       & 13           & Weapons     & 2            \\
Sexual Abuses      & 13           & Private Keys & 2            \\
Hacking            & 12           & Hitmen      & 1            \\
Counterfeit Bills & 7            & Citizenship & -            \\ \bottomrule
\end{tabular}%
\end{table}
\vspace{-0.2cm}

\subsection{Linking Activities on the Surface Web}
\label{sec:surfaceweb-activity} 
Another key characteristic of the campaigns we have identified is that they may extend their business networks beyond the dark web into the surface web. Despite such business owners can operate on both sides of the web in parallel, we have unveiled the previously neglected links, i.e., blockchain addresses, between them. 

By searching the $1,190$ illicit addresses using Google API, we have identified $13$ of them on the surface web. They are distributed across 44 URLs as listed in Table~\ref{tab:surface-website}. From the $13$ addresses, we trace back to the onion sites that belong to the 4 campaigns. In other words, 4 out of the 66 campaigns operate illicit sites both on the dark and the surface web. 

We further check whether there are differences in their activities across the two networks, and find two such URLs associated with one campaign. Table~\ref{tab:different-url} lists their details. By searching the term ``Cave Tor'' that appears in a page title, we manage to locate a dark web marketplace with the same name. Although the forum may not offer any illegal services or goods, it is still a part of the campaign's business network. This campaign is ranked as the most profitable one in Section~\ref{profitable-cluster}. Combined with the findings in Section~\ref{sec:speculator}, we confirm that the main body of this campaign is a marketplace called ``Cave Tor'' that profits from the sale of a variety of illicit goods.

\begin{table*}[t]
\centering
\caption{The campaign (identified by Bitcoin address) running various types of illicit sites across the surface and the dark web.}
\label{tab:different-url}
\resizebox{\linewidth}{!}{%
\begin{tabular}{@{}llll@{}}
\toprule
Page Title                                     & URL Category & BTC Address          & Onion Category \\ \midrule
Counterfeit money for sale — Dark social network &
  Counterfeit Bills &
  \multicolumn{1}{c}{\multirow{2}{*}{15BiCbHPscR6VXnJKkRg2W6UeEFCsjRGjs}} &
  \multirow{2}{*}{Cloned Card} \\
Registration — Dark social network - Cave Tor & Forums       & \multicolumn{1}{c}{} &                \\ \bottomrule
\end{tabular}%
}
\end{table*}

\subsection{Abuse Reports}\label{subsec:abuse_rep}
Considering that the illicit addresses and their onion sites may be complained by the surface web users, we particularly consider the abuse report to capture richer characteristics of the campaigns. In particular, we collect $47$ abuse reports according to Section~\ref{sec:study-design}. A large proportion of them is related to a variety of scams and frauds, complementing the categories included in our classification. We discuss several prominent examples in detail. 

\paragraph{A Campaign Involving Sexual Abuse} The address ``1Gs7Aztizk2\-rNNSE6AbpK4K7yAFTCZKV9a'', listed as the third profitable address in Section~\ref{sec:profiable-addr}, shows up six times in abuse reports~(i.e.,one court file~\cite{casetext} and five database records). They confirm its relevance to child pornography crimes where this address has been used as the payment method for several illicit onion sites. 

\paragraph{A Scam Campaign Using Stolen Onions} 
The abuse reports identify illicit campaigns using ``novel'' techniques. For example, 
we find a case of a hacker committing fraud using the onion sites and addresses stolen from benign dark web users. For example, the address ``1KRkAWDH5q7U5rdTM1\-rREmepk1pxpCAVKE'' found in $8$ categories of illegal onion sites (in Section~\ref{sec:overview}) is stolen from a benign user, according to a post on Twitter~\cite{cheenatwitter}. The user provides a screenshot of the stolen onion site with the domain name ``cheendv55ncemdot\-.onion". The screenshot shows the site is related to a fraud Western Union money transfer service. 

\paragraph{A Scam Campaign with Diversified Illicit Activities} 
The abuse reports help identify more illicit activities, complementing our clustering approach. 
For example, the address ``112FWGSL2q7rVTgabQ\-uJbo3WwKid8dMEtj'' is reported as a scam address according to five individual reports~(i.e., $4$ distinct databases and $1$ article~\cite{krishnan22medium}). The reports state that this address serves as the payment method for multiple dark web scam sites which provide services or goods such as hacker-for-hire, bank login credentials, and cloned cards. In comparison, this address is only associated with two Bitcoin generator scam sites and two account hacking service sites, based on our crawled dataset.

\paragraph{A Scam Campaign Advertising Anti-scam Services} 
The abuse reports also help identify addresses related to illicit campaigns that are not included in our dataset. 
For example, some interesting reports have mentioned that scammers can even advertise to provide public interest services such as anti-scam services. They trick the victims into donating to addresses that are actually controlled by themselves. 
Based on the reports, we identified one fake donation address ``1QATskw4LGVjhfB5UPZwiyVLKP9zdPcKir'', which has not been included in our dataset.

\vspace{0.1in}
\noindent\fbox{
	\parbox{0.95\linewidth}{
		\textbf{Answer to RQ3:} 
		We find most (69.7\%) of identified campaigns involve investment scams or cloned card sales for their high profitability. We identify four campaigns operating illicit sites on both the dark and surface web. Through their traces on the surface web, we reveal the entire business networks behind them. Furthermore, we have identified more diversified behaviors of illicit campaigns from 47 abuse reports. 
		 
}}
\vspace{0.1in}

\section{Discussions}\label{sec:discussions}
\subsection{Implications}
Our findings reveal the widespread cryptocurrency abuses in the dark web ecosystem. More specifically, a quarter of the collected onion sites contain blockchain addresses, out of which over 50\% are illicit. Their ubiquity sheds light on the necessity to develop an effective technique to detect and counteract such illicit activities. 
\subsubsection{Catching the Invisible Counterpart behind the Mirror} Even though service operators on the dark web can easily remain invisible against tracing and detection from law enforcement, their visible counterparts on the surface web could expose the entire business networks. This is based on our finding (see Section~\ref{sec:campaign}) that the dark web service operators may also run similar businesses on the surface web. Through blockchain address clustering, the hidden links between the visible and invisible parts can be unveiled. Therefore, this finding points out a promising direction for detecting illicit and suspicious dark web activities, leveraging the mapping towards the surface web using relevant blockchain address characteristics. 

\subsection{Limitations}\label{sec:limit}
In this work, we systematically study the status quo of cryptocurrency abuses among dark websites. However, it contains a few limitations that should be further investigated in future works.

\subsubsection{Dataset Comprehensiveness} As mentioned in Section~\ref{sec:study-design}, we have resorted to two data sources for identifying the onion addresses, which should comprehensively cover the high-profile sites recently accessible on the dark web. Nevertheless, we note that our searching approach is unable to discover all the accessible addresses in the wild. This is because onion addresses are intrinsically sequences of random characters, which renders them almost impossible to be discovered using traditional techniques (e.g., keyword searching), unless their owners publish them at some venues (e.g., surface websites or dark web indexing services). However, this limitation is unlikely to affect the fidelity of our findings due to the low popularity and potential impacts from the unidentified onion sites. 

\subsubsection{False Negatives from TF-IDF Classifier}
Our lightweight TF-IDF Classifier inevitably generates incorrect predictions, including false positives/negatives. The false positives can be effectively eliminated by checking the embedded blockchain addresses. The absence of illicit blockchain addresses deems the onion site to be false positive. However, the false negatives are more challenging to be removed precisely. In this study, we demonstrate the low false-negative rate of the classifier by manually checking the randomly sampled 100 unlabeled onion sites, as detailed in Section~\ref{subsec:classification_results}. 
\subsection{Ethical Considerations}
Similar to prior research works~\cite{lee2019cybercriminal,kanemura2019identification}, we took measures to avoid any potential legal disputes, conforming with the guidance of our government agency. Specifically, we refrained from tracking any information that could personally identify individuals and we were required to seek approval before sharing any information. To ensure that our research was conducted ethically and responsibly, we observe the following principles: (i) only collecting publicly accessible data, (ii) not tracking any personally identifiable information, (iii) restricting the access of textual data to the authors during the experimental period of this work, and (iv) releasing data under the supervision of the agency. We aim to unveil the previously unknown illicit campaigns on the dark web, minimizing their potential societal and financial impacts. 

\section{Related Works}\label{sec:related_works} 
In this section, we review the efforts in the literature investigating dark web crimes and cryptocurrency abuses. 
\subsection{Crimes on the Dark Web} 
As crimes are prevalent on the dark web, many researchers have studied and characterized the scale and impacts of them~\cite{vogt2017digital,liggett2020dark,baravalle2016mining,becker2018search,paoli2017behind,copeland2020assessing,anjum2021mysterious,wang2022crime,ogbanufe2023towards}. For example, Copeland et al.~\cite{copeland2020assessing} explored the illegal sale of firearms on the dark web while Sophia~\cite{vogt2017digital} discuss how to combat dark web crime on the legal framework. 
In addition to the trafficking of illegal goods, scams are prevalent on the dark web. One of them is the phishing scam, where scammers create sites with similar onion names and page content similar to well-known services and lure victims in. Since visitors cannot verify the authentication of an anonymous service as they do on the surface web, Yoon et al.~\cite{yoon2019doppelgangers} propose a method to identify phishing sites on the dark web based on content-duplicates and other external clues. Barr-Smith et al.~\cite{barr2020phishing} performed an analysis on ``typosquatting'' clone sites of onion services. 
Besides, some studies focus on the collecting and classification of dark web sites~\cite{ghosh2017automated,hayes2018framework,He2019ICISS,rawat2021dark,christin2022measuring}. For example, Ghost et al.~\cite{ghosh2017automated} proposed a classification framework based on manually selected keywords to automatically classify onion sites. We propose a comprehensive half-automated classification method of the dark web based on TD-IDF and page similarity measurement in Section~\ref{sec:classify}. 

The most related research topic to our work in the dark web crime study is analyzing illicit blockchain addresses on the dark web~\cite{lee2019cybercriminal,kanemura2019identification,dearden2023follow}. Lee et al.~\cite{lee2019cybercriminal} take the first step toward exposing illicit activities involving the dark web and cryptocurrency. They implement a platform that collects dark websites and extracts useful cryptocurrency information from them automatically. They also provide diverse case studies of cryptocurrency usage on the dark web. Kanemura et al.~\cite{kanemura2019identification} proposed a machine learning approach to identify Bitcoin addresses in the dark markets. They extracted 73 features of Bitcoin addresses to train the supervised classifiers and further applied simple multi-input heuristics to expand the addresses they identified. Dearden et al~\cite{dearden2023follow} focused exclusively on analyzing transactions from the AlphaBay darknet marketplace. They conducted a quantitative analysis using the Bitcoin blockchain ledger, examining both legal and illegal cryptocurrency activities. However, the studies above do not study the address campaign in the dark web or just use simple heuristics for address expansion. In this work, we take efforts to identify and characterize cryptocurrency address campaigns controlled by dark web attacker groups through a comprehensive approach, as detailed in Section~\ref{sec:cluster} and Section~\ref{sec:campaign}. 

\subsection{Cryptocurrency Abuses} 
Besides some studies that target cryptocurrency crimes like money laundering~\cite{barone2019cryptocurrency,reddy2018cryptocurrency,comolli2021surfing,pettersson2022combating} and ransomware~\cite{kshetri2017crypto,leverett2020averages,alqahtani2022survey,kok2022early,kotov2023understanding}, many studies on cryptocurrency abuses are focused on cryptocurrency scams. Although users, wallets and exchanges are adopting new countermeasures to avoid being scammed, new scam techniques still emerge to defraud users' money. Vasek and Moore surveyed the presence of Bitcoin-based scams~\cite{vasek2015there} in 2015. By gathering reports from voluntary vigilantes and reports tracked in online forums, they identified 192 scams and categorized them into four groups: \textit{Ponzi schemes}, \textit{mining scams}, \textit{scam wallets} and \textit{fraudulent exchanges}. 
Other existing abuse studies were focused on detecting the Ponzi schemes~\cite{chen2018detecting, bartoletti2020dissecting, bartoletti2018data,vasek2018analyzing,toyoda2019novel,zhang2021detecting,jin2022heterogeneous,heyman2023red}, fake/scam tokens~\cite{gao2020tracking,xia2021trade,mazorra2022not,imeri2023smart}, market manipulation of cryptocurrencies~\cite{gandal2018price,chen2019market,chen2019detecting,nizzoli2020charting,mirtaheri2021identifying,kampers2022manipulation,dhawan2023new}, blockchain honeypots~\cite{torres2019art,hara2021machine,hongxia2022honeypot}, and phishing scams~\cite{wu2019phishers,phillips2020tracing,wang2021tsgn,fu2022ct,wen2023novel}. 

\section{Conclusion} \label{sec:conlcusion} 
In this work, we carry out a systematic study tracking the cryptocurrency abuses related to illicit campaigns in the dark web ecosystem. We collect and construct a dataset containing around 5K dark websites and over 130K pages. Utilizing this dataset, we build a classifier to detect the illicit sites, and extract the blockchain addresses to further characterize the cryptocurrency abuses. From our study, we have identified 2,564 illicit sites with 405 blockchain addresses, accounting for over \$2.64 million in transaction volume. Furthermore, we have identified 66 illicit campaigns behind 1,396 onion sites, utilizing their underlying connections. Our work intends to raise awareness of the prevalent and persistent illicit activities and cryptocurrency abuses on the dark web. Meanwhile, our study should also facilitate and encourage future work in the area of illicit dark website detection based on blockchain addresses.

\balance
\bibliography{ref}

\section*{Declarations} 

\noindent \textbf{Ethical Approval} Not applicable since there are no human and/or animal studies included in this paper. 

\noindent \textbf{Competing Interests} The authors have no competing interests to declare that are relevant to the content of this article. 

\noindent \textbf{Funding} There is no external funding received for this work. 

\noindent \textbf{Availability of data and materials} To facilitate future research, we will responsibly release our dataset to trusted entities on a per-request basis.

\end{document}